\documentclass[twocolumn,secnumarabic,amssymb, nobibnotes, aps, prl, superscriptaddress, nobalancelastpage]{revtex4-1}

\usepackage{graphicx}
\usepackage{amsmath}
\usepackage{epsfig}
\usepackage{graphicx}
\usepackage{braket}
\usepackage{amssymb,tabularx,dcolumn}
\usepackage{supertabular,ltxtable}
\usepackage{longtable}
\usepackage{float}
\usepackage{color}
\usepackage{bm}
\usepackage{hyperref}
\usepackage[free-standing-units]{siunitx}
\usepackage[useregional]{datetime2}
\newcommand{\oSix}{\ensuremath{^{16}}O}

\newcommand{\oEight}{\ensuremath{^{18}}O}
\newcommand{\oSixEight}{\ensuremath{^{16,18}}O}

\newcommand{\neEight}{\ensuremath{^{18}}N\lowercase{e}}

\newcommand{\caForty}{\ensuremath{^{40}}C\lowercase{a}}
\newcommand{\caEight}{\ensuremath{^{48}}C\lowercase{a}}
\newcommand{\caAughtEight}{\ensuremath{^{40,48}}C\lowercase{a}}

\newcommand{\niSix}{\ensuremath{^{56}}N\lowercase{i}}
\newcommand{\niEight}{\ensuremath{^{58}}N\lowercase{i}}

\newcommand{\niFour}{\ensuremath{^{64}}N\lowercase{i}}
\newcommand{\niEightFour}{\ensuremath{^{58,64}}N\lowercase{i}}

\newcommand{\snHundred}{\ensuremath{^{100}}S\lowercase{n}}
\newcommand{\snTwelve}{\ensuremath{^{112}}S\lowercase{n}}

\newcommand{\snFour}{\ensuremath{^{124}}S\lowercase{n}}
\newcommand{\snTwelveFour}{\ensuremath{^{112,124}}S\lowercase{n}}

\newcommand{\pbEight}{\ensuremath{^{208}}P\lowercase{b}}

\newcommand{\sOne}{s\ensuremath{_{1/2}}}
\newcommand{\pThree}{p\ensuremath{_{3/2}}}

\newcommand{\dFive}{d\ensuremath{_{5/2}}}

\newcommand{\fSeven}{f\ensuremath{_{7/2}}}

\begin{document}

\title{Systematic Matter and Binding-Energy Distributions\\
from a Dispersive Optical Model Analysis}

\author{C.D.~Pruitt}%
\email[Corresponding author: ]{pruitt9@llnl.gov}
\altaffiliation{Present Address: \textit{Lawrence Livermore
National Laboratory, Livermore, CA}}
\affiliation{Department of Chemistry, Washington University, St. Louis, MO 63130, USA}

\author{R.J.~Charity}
\affiliation{Department of  Chemistry, Washington University, St. Louis, MO 63130, USA}

\author{L.G.~Sobotka}
\affiliation{Department of Chemistry, Washington University, St. Louis, MO 63130, USA}
\affiliation{Department of Physics, Washington University, St. Louis, MO 63130, USA}

\author{M.C.~Atkinson}
\altaffiliation{Present Address: \textit{TRIUMF, Vancouver, Canada}}
\affiliation{Department of Physics, Washington University, St. Louis, MO 63130, USA}

\author{W.H.~Dickhoff}
\affiliation{Department of Physics, Washington University, St. Louis, MO 63130, USA}

\begin{abstract}
    We present the first systematic nonlocal dispersive-optical-model analysis using both
    bound-state and scattering data of \oSixEight, \caAughtEight, \niEightFour, \snTwelveFour,
    and \pbEight. In all systems, roughly half the total nuclear binding energy is associated with the
    most-bound 10\% of the total nucleon density. The extracted neutron skins
    reveal interplay of asymmetry, Coulomb, and shell effects on the skin
    thickness. Our results indicate that simultaneous optical-model fits of inelastic
    scattering and structural data on isotopic pairs are effective for
    constraining asymmetry-dependent nuclear structural quantities otherwise difficult
    to observe experimentally.
\end{abstract} 

\maketitle

\textit{Introduction.--}
Despite much investigation, the detailed behavior of individual nucleons in
the nuclear ground state remains poorly understood. While many models can 
reproduce nuclear masses and charge radii across the nuclear chart, none
can fully account for the \textit{distribution} -- in radial,
energy, momentum, and angular-momentum space -- of nucleons within the nuclear volume.
For example, the precise location of excess neutrons in neutron-rich systems like \caEight\ and
\pbEight\ remains an open question, one that has received immense theoretical and experimental interest in
recent years (thoroughly reviewed in \cite{Thiel2019}). The existence of ``neutron skins'', $\Delta r_{np}$, on the surface of these
and other stable nuclei is expected to correlate strongly with the density dependence of the
nuclear symmetry energy, a major uncertainty in the neutron-star equation-of-state
\cite{Fattoyev2012, Piekarewicz2012, Vinas2014}. Experimental
difficulties in direct neutron-skin measurements and uncertainty about the sensitivity of mean-field
models to isovector quantities \cite{Furnstahl2002} make alternative
approaches desirable. Ideally, a comprehensive model should not only reproduce
integrated quantities (like the charge radius or total binding energy)
but also specify how nucleons share momentum and energy, all
while being realistic about the model uncertainty of its predictions
\cite{UncertaintyEditorial}.

A step toward these goals was the establishment of the
dispersive optical model (DOM) \cite{Mahaux1991}, which formally extended
traditional optical potentials to negative energies so that both reaction and
structural information could be used to probe the nuclear
potential (see the reviews of \cite{Dickhoff2017, Dickhoff2019}). Previous DOM case studies
have shown promise for exploring systematics of nucleon behavior:
for instance, generating trends in valence-shell spectroscopic factors as a function of asymmetry
\cite{Mueller2011, Atkinson2019} and momentum \cite{Atkinson2018}, and extracting neutron
skins \cite{Mahzoon2014, Mahzoon2017, Atkinson2020}.
However, each of these studies was narrow in scope: \cite{Mahzoon2014,
Mahzoon2017, Atkinson2018} examined only a single Ca isotope each, while 
\cite{Mueller2011} included almost no bound-state information and thus
was mute about matter and energy distributions.
Cognizant of these challenges, we have completed a joint DOM analysis of the doubly-
and single-closed-shell nuclei \oSixEight, \caAughtEight,
\niEightFour, \snTwelveFour, and \pbEight, the
first multi-nucleus treatment of nucleon matter and binding energy distributions
in an optical-model framework. For parameter optimization and uncertainty characterization,
we relied on Markov Chain Monte Carlo (MCMC) sampling, an important improvement over techniques used
for previous state-of-the-art optical potentials \cite{CH89,KoningDelaroche}.
In all nine isotopes we examined, the small fraction of
nucleon density far below the deepest single-particle
energies (e.g., below -100 MeV) was found to play a critical role for reproducing experimental
binding energies. Before presenting these results, we first review salient elements of the DOM
formalism.

\textit{Relevant DOM Formalism.--}
Classical optical models describe nucleon-nucleus scattering with various forms of local phenomenological
potentials defined only at positive energies \cite{BecchettiGreenlees, CH89, KoningDelaroche}. In
contrast, the DOM defines the complex, nucleon self-energy (or effective interaction), $\Sigma^{*}(\alpha,\beta;E)$, both above and below the
Fermi energy. This potential-like object dictates
 nucleon behavior as it moves from state $\alpha$ to state $\beta$ in the nuclear
medium at energy $E$, where $\alpha,\beta$ denote complete sets of
quantum numbers sufficient to specify the single-particle state.
As in past DOM treatments \cite{Mahzoon2014, Atkinson2018, Atkinson2019,
Atkinson2020}, the self-energy domain was restricted
to $-300$ MeV to $200$ MeV with respect to the Fermi energy, a first-order
relativistic correction was included, and only two-body forces were considered.
The self-energy is comprised of three subcomponents:
\begin{equation} \label{SelfEnergyBreakdown}
        \Sigma^{*}(\alpha,\beta;E) = \Sigma_{s}(\alpha,\beta) + \Sigma_{im}(\alpha,\beta;E)
        + \Sigma_{d}(\alpha,\beta;E)
\end{equation}
The ``static'' part of the self-energy $\Sigma_{s}(\alpha,\beta)$ includes all
real energy-independent contributions, taken here as a Hartree-Fock term
evaluated at the Fermi energy, plus a spin-orbit term. The Fermi energy is defined via the
ground-state energies for the $A\pm1$ systems:
\begin{equation} \label{FermiEnergyDefinition}
    \epsilon_{F} \equiv \frac{1}{2}[E^{A+1}_{0}+E^{A-1}_{0}].
\end{equation}
Each of the real subterms are parameterized with a Woods-Saxon form (or its derivative) coupled
to a Gaussian nonlocality. The energy-dependent imaginary component
$\Sigma_{im}(\alpha,\beta;E)$ consists of energy-dependent surface- and volume-associated
terms at both positive and negative energies, again with nonlocal Woods-Saxons,
or their derivatives, for radial dependence. Physically, these terms account for
inelastic processes that require the most computational effort to recover
in \textit{ab initio} and shell-model treatments.
To constrain these terms, the DOM instead relies on fitting flexible potential forms to
experimental data. The ``dynamic'' (energy-dependent) real term $\Sigma_{d}(\alpha,\beta;E)$ is
completely determined by integrating the imaginary term over the entire energy
domain. It ensures that the self-energy obeys the required subtracted dispersion relation.
The parameterization used is available in the companion article
\cite{Pruitt2020PRC}; additional detail can be found in \cite{PruittPhDThesis}.

Following \cite{MBTE}, the single-nucleon propagator is generated from the
self-energy via the Dyson Equation and the hole spectral function
extracted from the propagator:
\begin{equation}
    \begin{split}
        S_{\ell j}^{h}(\alpha; E) & =
        \frac{1}{\pi}\operatorname{Im}({G_{\ell j}(\alpha,\alpha;E)})\qquad \text{for }
        E\leq\epsilon_{F}.
    \end{split}
\end{equation}
Here $G$ and $S$ are labeled with the (conserved) 
angular momentum $\ell$ and total angular momentum $j$. Intuitively, the hole spectral function
is the probability for removal of a particle with quantum numbers $\alpha$ from an initial $A$-body system
with ground-state energy $E^{A}_{0}$, leaving the residual ($A$-1)-body system with remaining energy
$E^{A}_{0}-E$. Taking an explicit $r$-space basis for $\alpha$, the nucleon point density can be directly
calculated from the hole spectral function:
\begin{equation} \label{PointDensityEquation}
    \rho_{\ell j}(r) = \frac{1}{4\pi r^{2}} \int_{-\infty}^{\epsilon_{F}} (2j+1) S_{\ell j}^{h}(r; E)
dE.
\end{equation}

The total binding energy can be calculated per the Migdal-Galitsky rule, which is exact
when only two-body interactions are included:
\begin{equation} \label{MigdalGalitsky}
    \begin{split}
        E^{A}_{0} & = \frac{1}{2} \left[\sum_{\alpha\beta}\braket{\alpha|\hat{T}|\beta}n_{\alpha,\beta}
        + \sum_{\alpha}\int_{-\infty}^{\epsilon_{F}}dE\ E\ S_{\ell j}^{h}(\alpha;E)\right]
    \end{split},
\end{equation}
where $\hat{T}$ is the kinetic energy operator appropriate for the
single-particle basis and $n_{\alpha,\beta}$ is the one-body density matrix.
Three-body terms do not induce important corrections when energy densities
are considered, supporting the use of Eq.(\ref{MigdalGalitsky}) in DOM applications
\cite{Atkinson2020_2}.

To constrain the self-energy, we applied nine sectors of experimental data for each nucleus: differential
elastic-scattering cross sections, analyzing powers, reaction cross sections, total cross sections,
binding energies, charge radii, charge densities,
single-nucleon separation energies, and particle numbers. For fits on \oSixEight,
\caAughtEight, \niEightFour, and \snTwelveFour, all available data for each isotope pair were
\textit{simultaneously} fit using the same asymmetry-dependent potential; for \pbEight, only the \pbEight\ data
were used. The new experimental isotopically-resolved neutron total
cross sections that motivated this work are reported in the companion experimental
paper \cite{Pruitt2020PRC}, which also includes detailed comparison of
DOM calculations to all experimental data, specifics of the MCMC implementation, and parameter
estimates with uncertainties.

\textit{Binding Energies.--}
\begin{figure*}[!htb]
    \includegraphics[width=\linewidth]{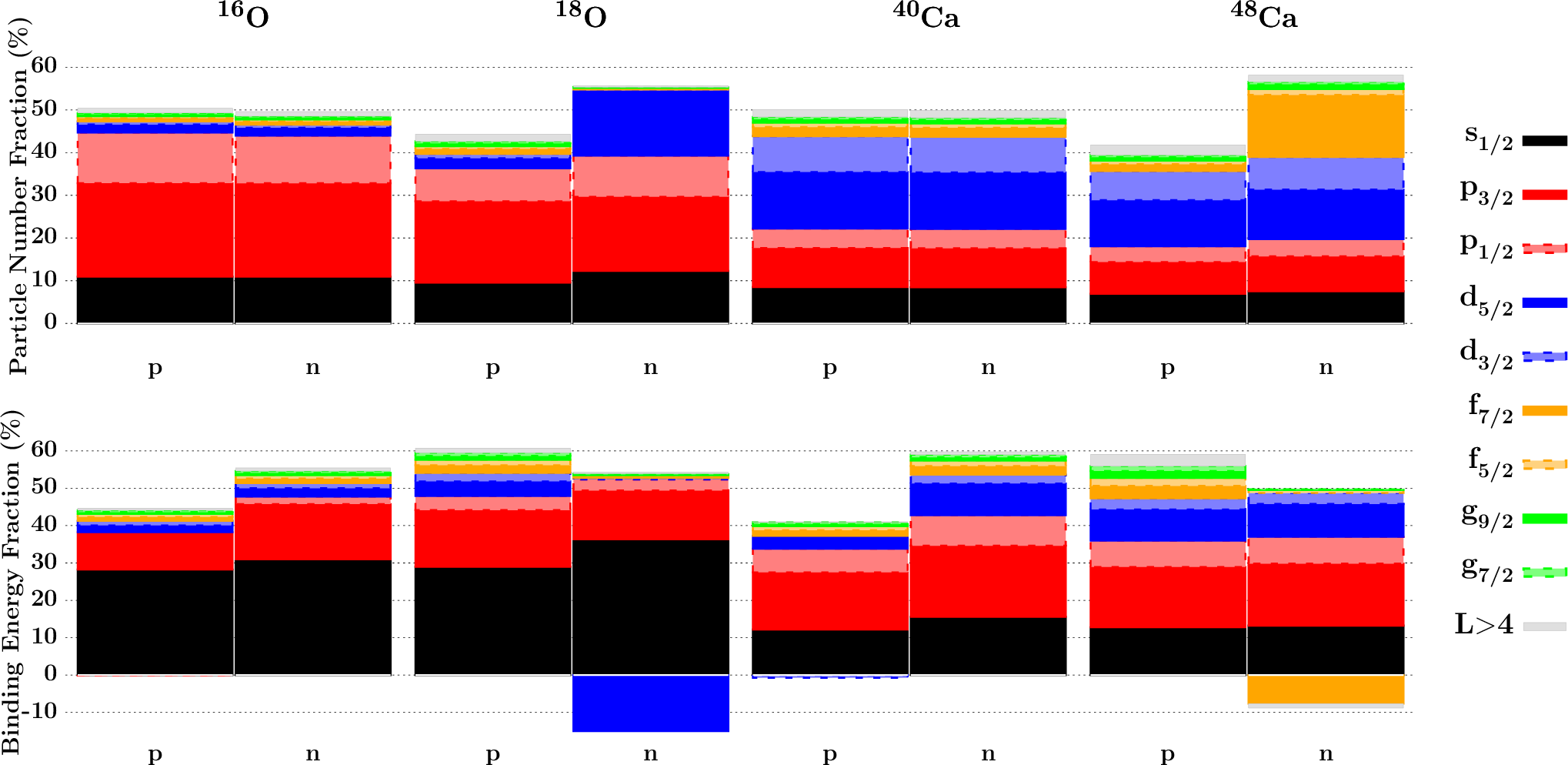}
    \caption{\label{OccupationAndBinding}DOM calculations of nucleon occupation
        and binding energy contributions as a function of angular
        momenta $\ell j$ in \oSixEight\ and \caAughtEight. The results shown are
        using the median posterior parameter values from MCMC sampling.}
\end{figure*}
\begin{figure}[!htb]
    \includegraphics[width=\linewidth]{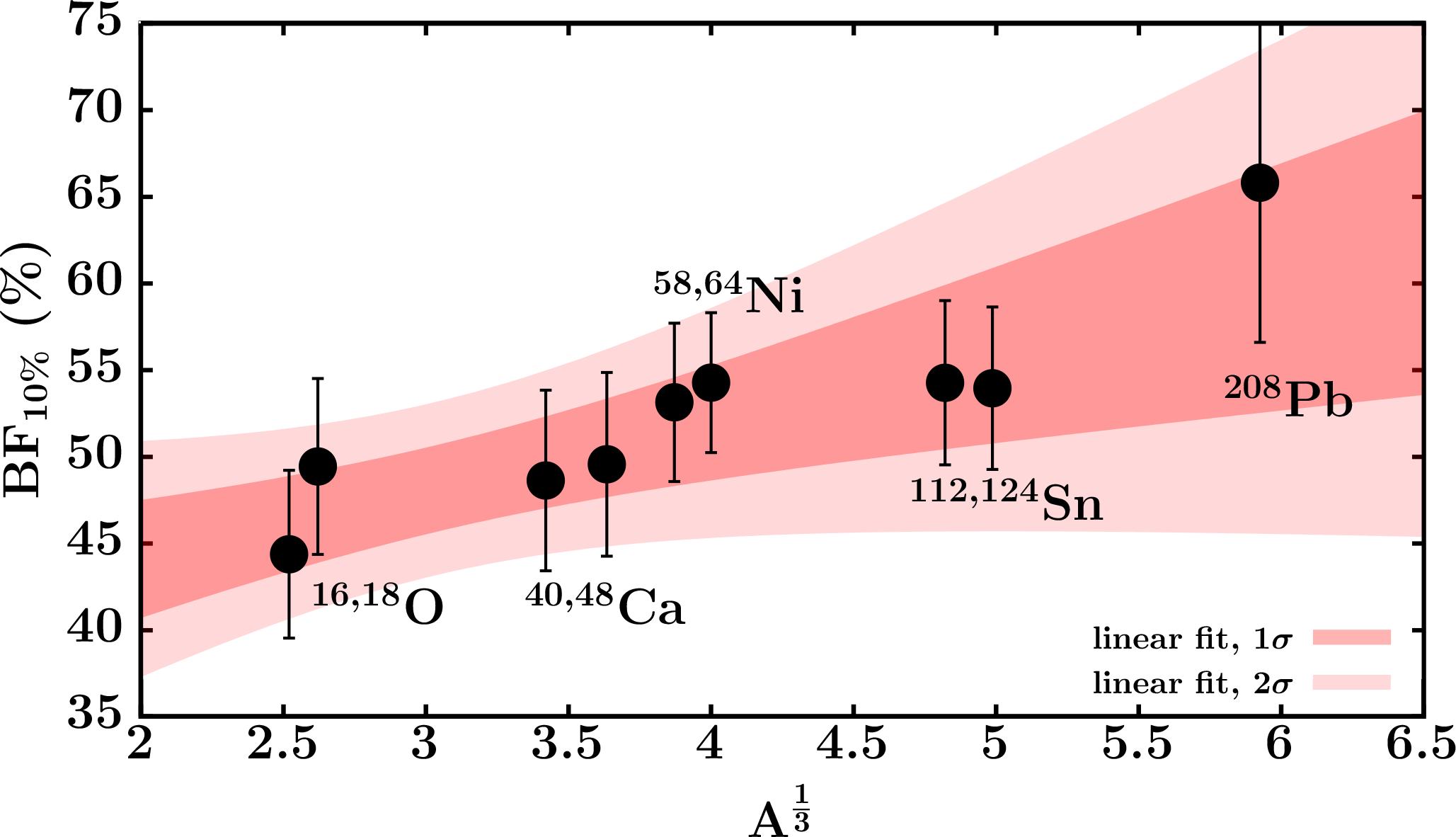}
    \caption{\label{BerniePlot} Fraction of the total binding energy possessed by the
        most-deeply-bound 10\% of the nucleon density for the isotopes
        studied in this work. The shaded regions indicate parametric uncertainty
        from fitting Eq. \ref{BernieFractionModel} to these data.}
\end{figure}
Figure \ref{OccupationAndBinding} shows the breakdown of particle density and binding
energy for optimized fits of \oSixEight\ and \caAughtEight. As in an independent-particle model, the vast majority of
both proton and neutron density rests in the shells below the Fermi level. However, due to the
imaginary potentials, a significant fraction, around 10\%, appears in higher shells
that would be fully unoccupied in a na\"ive mean-field picture.

For both protons and neutrons, an outsized fraction
of the binding energies comes from the most-bound levels. For example, the \sOne\ states in
\oSix\ possess roughly 20\% of the total nucleon density but almost 60\% of the total binding energy.
This is a consequence of the long tail of the hole spectral functions extending to extremely negative energies
(more than 100 MeV below $\epsilon_{F}$), far below the mean-field expectation.
In both systems, the protons' fraction of the total binding energy is
slightly reduced (less bound) compared to that from the neutrons, a consequence of the Coulomb interaction.
Overall, the substantial depletion of
mean-field occupancies even in light systems (and associated broadening of the bound-nucleon spectral
functions, as illustrated in \cite{Atkinson2020}) is critical for achieving an average
binding energy of 8 MeV/nucleon. We note that the binding energy distribution among
shells that we recover for \oSix\ agrees with that from the Brueckner-Hartree-Fock (BHF) treatment of \cite{Muther1995}
and with general features of \textit{ab initio} many-body calculations for nuclear matter \cite{Vonderfecht1993}.

Finally we turn to the binding energy distributions for asymmetric \oEight\ and
\caEight\ in Fig. \ref{OccupationAndBinding}. In these systems, the minority species (protons) experiences
a deeper mean-field potential and a larger imaginary potential, increasing each proton's relative share of
the binding energy. For the majority species (neutrons), the
effect is reversed: binding is reduced (less bound) for each shell compared to the symmetric
system. For the valence \dFive\ neutrons in \oEight\ (in blue) and \fSeven\ neutrons in \caEight\
(in orange), the contribution to the total binding is negative --- that is,
\textit{unbinding} -- because the bulk of their spectral density resides in quasiholes
at or near the Fermi surface. This effect is more than compensated by the extra
binding energy these valence neutrons induce in the \textit{protons} compared to the symmetric case,
such that the net effect is increased overall binding. These results are consistent with enhancement
of short-range correlations (SRCs) among minority nucleons as identified by \cite{Hen2012} in their
investigation of nucleon high-momentum content as a function of asymmetry.

\begin{figure*}[!htb]
    \includegraphics[width=0.8\linewidth]{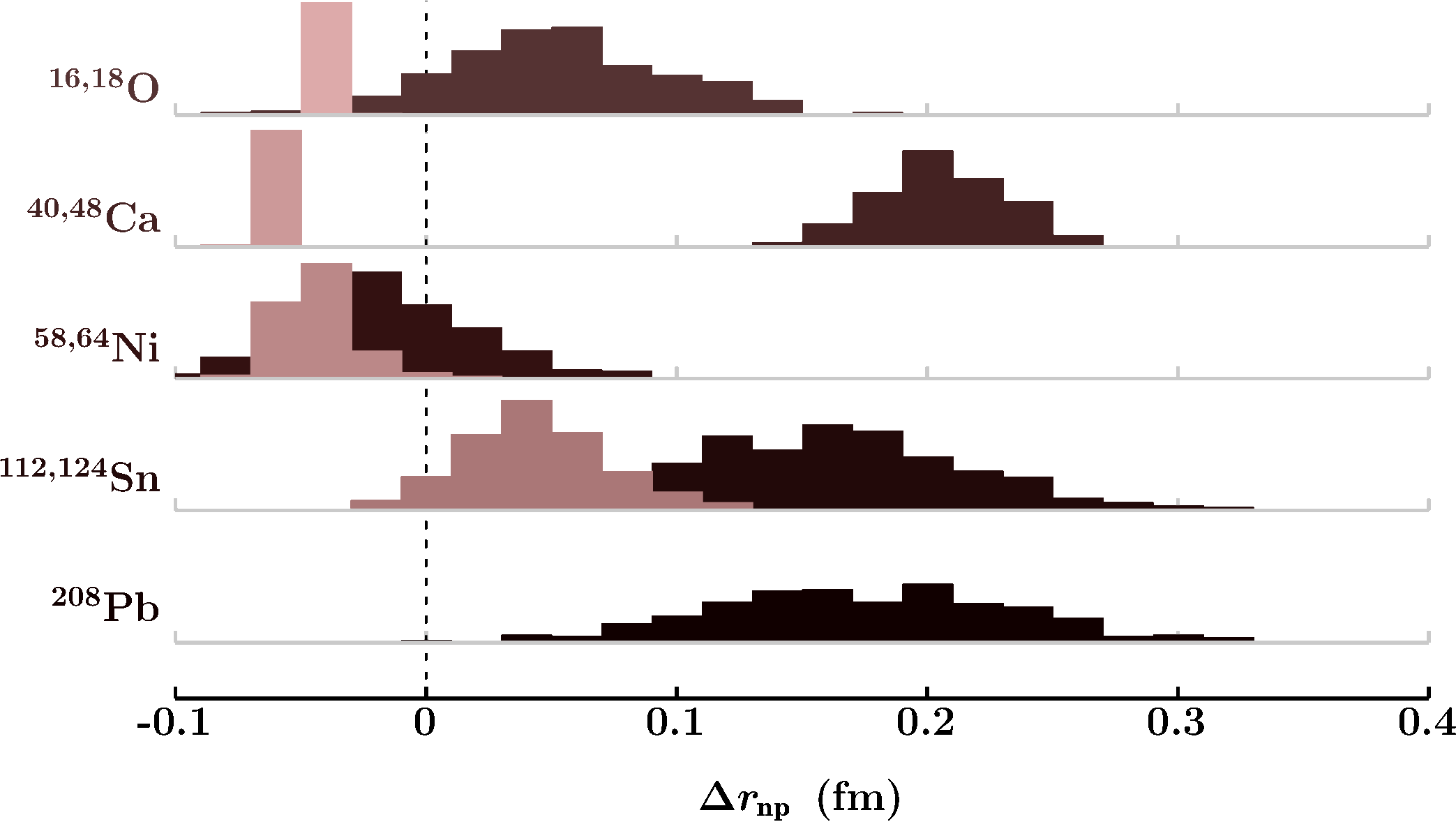}
    \caption{\label{skinsPlot} Neutron-skin probabilities via MCMC sampling for
        \oSixEight, \caAughtEight, \niEightFour, \snTwelveFour, and \pbEight.
        Each axis shows a single element. For elements with two isotopes
        histogrammed, the lighter isotope is shown using light bars,
    and the heavier isotope is shown with dark bars. The heights of each 
    distribution have been arbitrarily rescaled to facilitate comparison.}
\end{figure*}

Figure \ref{BerniePlot} gives an $\ell j$-independent illustration of systematic behavior of the binding
energy distribution. For each system, the fraction of the total binding energy possessed
by the most-bound 10\% of the total nucleon density ($BF_{10\%}$), regardless of 
quantum number, is plotted. The error bars indicate the 16\textsuperscript{th} and 84\textsuperscript{th} percentiles
from the MCMC posterior distributions (the 1$\sigma$-region, if the posteriors are assumed to
be Gaussian). For all systems analyzed here, $BF_{10\%}$ exceeds 40\%.
To put this percentage in context, we performed an analogous ``single-particle'' calculation
on \caForty\ by artificially placing all spectral density for the \sOne\ nucleons at their lowest single-particle eigenvalue.
This scenario yields a $BF_{10\%}$ of 31\% for \caForty, much lower than the median value of
48\% from Fig. \ref{BerniePlot}, demonstrating that the tiny nucleon density at extremely negative (deeply bound)
energies makes an outsized contribution to overall binding.

To determine the relative effect of nuclear size and asymmetry on this
quantity, we applied a linear model to the data,
\begin{equation}\label{BernieFractionModel}
    BF_{10\%} = x_{0} + A^{\frac{1}{3}} x_{A} + \frac{N-Z}{A}
        x_{\alpha},
\end{equation}
with $N$, $Z$, and $A$ the neutron, proton, and total nucleon numbers. MCMC sampling of this
model gives parameter posterior values of $x_{0} = 36^{44}_{30}$, $x_{A} =
4.1^{6.1}_{1.5}$, and $x_{\alpha} = 3^{31}_{-26}$, where the 16\textsuperscript{th},
50\textsuperscript{th}, and 84\textsuperscript{th} percentile values are reported as $\text{50}^{\text{84}}_{\text{16}}$.
Thus the $BF_{10\%}$ depends only weakly on
the size of the system and is independent of asymmetry, indicating
that even in light nuclei, the bulk of the total binding comes from the few most-bound nucleons.

\textit{Neutron skins.--}
The neutron skin:
\begin{equation}
    \Delta r_{np} \equiv r_{rms}(n) - r_{rms}(p),
\end{equation}
was first identified as an important observable
by Wilkinson over fifty years ago \cite{Wilkinson1967}. Neutron skins on neutron-rich
nuclei are connected to other nuclear structural quantities,
including the electric dipole polarizability, the location of the pygmy and giant
dipole resonances, the density dependence of the symmetry energy,
and the size of neutron stars \cite{Vinas2014, Brown2000, Fattoyev2012, Zhang2018, TypelBrown2001}.

The neutron skins extracted from the present work
are shown in Fig. \ref{skinsPlot} and median values and uncertainties in Table \ref{SkinsTable}.
We find that the degree of asymmetry, $\alpha \equiv \frac{N-Z}{A}$,
correlates strongly ($r=0.89$) with the median skin thicknesses.
If a simple linear dependence in $\alpha$ is assumed, extrapolation from the
\niEightFour\ skins gives a \niSix\ skin thickness of -0.04 $\pm$0.03 fm.
A similar calculation with \snTwelveFour\ yields a \snHundred\ skin thickness of -0.07 $\pm$0.06 fm.
In the symmetric systems \oSix\ and \caForty, Coulomb repulsion nudges
proton density outward from the core, resulting in a small negative neutron skin (that is, a proton
skin). Again assuming linear dependence of this Coulomb effect,
extrapolation from \oSix\ and \caForty\
gives neutron skins of -0.07$\pm 0.02$ fm for \niSix\ and
-0.12$\pm 0.04$ fm for \snHundred, slightly more negative than, but in keeping with, the linear
extrapolation from \niEightFour\ and \snTwelveFour. Besides Coulomb and asymmetry-dependent effects,
the large \caEight\ median skin of $0.22$ fm and near-zero median \niFour\ skin of -0.01 fm show
the importance of shell effects for certain systems (cf. with \pbEight\ results of 
\cite{Atkinson2020}). To wit, most 
of the excess neutrons in \caEight\ and \niFour\ enter the neutron \fSeven\ and neutron \pThree\ shells,
respectively, as seen in Fig. \ref{OccupationAndBinding} for \caEight. The mean radius of
the $\fSeven$ shell is larger than the deeper shells; thus, when neutron density is added,
the size grows rapidly. In \niFour, the neutron 1\pThree\ rms radius is
closer to the overall $r_{rms}(n)$ of \niEight, so the additional neutrons of \niFour\ do little to grow
the skin thickness.

\begin{table}[tb]
    \centering
    \caption{
        Neutron skins ($\Delta r_{np}$), in fm, from this work. The 16\textsuperscript{th},
    50\textsuperscript{th}, and 84\textsuperscript{th} percentile values of the skin distribution
    are reported as $\text{50}^{\text{84}}_{\text{16}}$.
}
    \bgroup
    \def\arraystretch{1.8}%
    \begin{tabular}{c c c c}
        \oSix & \oEight  & \caForty & \caEight \\
        \hline
        -0.025$^{-0.023}_{-0.027}$ &
        0.06$^{0.11}_{0.02}$ & 
        -0.051$^{-0.048}_{-0.055}$ &
        0.22$^{0.24}_{0.19}$ \\
    \end{tabular}
    \begin{tabular}{c c c c}
        \niEight & \niFour & \snTwelve & \snFour \\
        \hline
        -0.03$^{-0.02}_{-0.05}$ &
        -0.01$^{0.03}_{-0.04}$ &
        0.05$^{0.08}_{0.02}$ &
        0.17$^{0.23}_{0.12}$ \\
    \end{tabular}
    \begin{tabular}{c}
        \pbEight \\
        \hline
        0.18$^{0.25}_{0.12}$ \\
    \end{tabular}
    \egroup
    \label{SkinsTable}
\end{table}

For \oEight, the mirror-nuclei logic of \cite{Brown2017} can be applied to cross-check
our skin value. Assuming isospin symmetry, the difference between
the \neEight\ and \oEight\ charge radii is a good proxy for the \oEight\ neutron skin thickness. Per \cite{Angeli2013},
the charge radius difference between \neEight\ and \oEight\ is 0.20$\pm$0.01 fm. Before comparing this
proxy value with the neutron skin of \oEight, Coulomb and deformation corrections must be applied.
First, due to the Coulomb force, the proton density of \neEight\ extends further than the neutron
density of \oEight. We estimate the magnitude of this proton density extension in \neEight\ as 0.03
fm, or 25\% larger than the difference between the proton and neutron distributions of \oSix,
due the 25\% larger proton number of \neEight. Subtracting 0.03 fm from the \neEight-\oEight\
radius difference yields 0.17 fm. Second, because \neEight\ is more deformed
($\beta_{2}=0.68$) than \oEight\ ($\beta_{2}=0.37$) \cite{NUDATDatabase}, any deformation correction
will further reduce this proxy value. Absent a clean way to generate this correction, the proxy can only be taken
as an \textit{upper limit} on the \oEight\ neutron skin. Our skin prediction for \oEight\ of
0.06$^{0.11}_{0.02}$ fm is compatible with the upper limit of 0.17 fm provided by this
heuristic symmetry argument.

Our median results for \caEight\ (0.22 fm) and \pbEight\
(0.18 fm) are somewhat smaller than those from previously-mentioned
DOM case studies but with significant uncertainty-range overlap \cite{Mahzoon2017, Atkinson2020}.
We attribute the variation to differences in the potential parameterization, our joint fitting of
isotope pairs, and our MCMC optimization approach. The values reported here for \caEight, \pbEight\ are
quite close both to those from recent experimental studies
\cite{Zenihiro2018, Tarbert2014, Zenihiro2010} and to those from the relativistic density functional model
FSUGold as reported in \cite{Thiel2019}. However, our predicted skin range for
\caEight\ differs significantly from the recent coupled-cluster-based prediction of 0.12-0.15 fm from \cite{Hagen2016},
a discrepancy we hope the proposed CREX experiment will resolve. Lastly, the median skins we recover
for \snTwelveFour\ (0.05 and 0.17 fm, respectively) are almost identical to those extracted by
\cite{Terashima2008} (0.06 and 0.18 fm, respectively) from analysis of
295-MeV proton elastic scattering on Sn isotopes. 

\textit{Conclusions.--}
Using a newly-generalized version of the DOM, we performed the first
systematic DOM analysis across nine isotopes from $A$=16 to $A$=208
to extract matter and binding-energy distributions. Using MCMC with
model discrepancy terms and joint fitting of multiple
isotopes, we generated realistic uncertainties for all potential parameters and extracted quantities.
Our results quantitatively indicate how asymmetry, Coulomb, and shell effects contribute to
neutron skin generation and drive a disproportionate share of
the total binding energy to the deepest nucleons.
Using simple trends in \oSix, \caForty, \niEightFour, and
\snTwelveFour, we estimate the \niSix\ neutron skin as between -0.04 and -0.07 fm
and between -0.07 and -0.12 fm for \snHundred. Our
skin thickness for \oEight\ agrees with the mirror-nucleus upper bound expectation,
and the agreement of our \caEight, \snTwelveFour, \pbEight\ skin thicknesses with recent external
predictions augers well for a future truly global DOM treatment.

\textit{Acknowledgments.--}
This work is supported by the U.S. Department of Energy, Office of Science, 
Office of Nuclear Physics under award numbers DE-FG02-87ER-40316,
by the U.S. National Science Foundation under grants PHY-1613362 and PHY-1912643,
and was performed in part under the auspices of the U.S. Department of Energy
by Lawrence Livermore National Laboratory under Contract DE-AC52-07NA27344.
C.D.P. acknowledges support from the Department of Energy, National Nuclear Security
Administration, under Award Number DE-NA0003841, the Center for Excellence in
Nuclear Training And University-based Research (CENTAUR).

\bibliographystyle{apsrev4-1}
\bibliography{references}

\end{document}